\documentstyle[12pt]{article}

\begin{document}
\hfuzz=10pt

\font\twelvemb=cmmib10 scaled \magstep1
\textfont9=\twelvemb
\def\boldtau{\fam9\mathchar"711C}

\newcommand{\fracsm}[2]{{\textstyle\frac{#1}{#2}}}
\newcommand{\ur}[1]{(\ref{#1})}
\newcommand{\eq}[1]{eq.~(\ref{#1})}
\newcommand{\eqs}[2]{eqs.~(\ref{#1},~\ref{#2})}
\newcommand{\eqsss}[2]{eqs.~(\ref{#1}--\ref{#2})}
\newcommand{\Eq}[1]{Eq.~(\ref{#1})}
\newcommand{\Eqs}[2]{Eqs.~(\ref{#1},~\ref{#2})}
\newcommand{\Eqsss}[2]{Eqs.~(\ref{#1}--\ref{#2})}
\newcommand{\fig}[1]{Fig.~\ref{#1}}
\newcommand{\beq}{\begin{equation}}
\newcommand{\eeq}{\end{equation}}
\newcommand{\la}[1]{\label{#1}}
\newcommand{\SU}{$SU(2)~$}
\newcommand{\e}{\varepsilon}
\newcommand{\doublet}[3]{\: \left(\begin{array}{c} #1 \\#2
            \end{array} \right)_{#3}}
\newcommand{\matr}[4]{\left(\begin{array}{cc}
#1 &#2 \\
#3 &#4                       \end{array} \right)}
\let\oldchi=\chi
\renewcommand{\chi}{\raise 2pt\hbox{$\oldchi$}}

\thispagestyle{empty}
\vspace{2cm}
\begin{center} {\Large\bf
Loop corrections to the sphaleron transition rate in the minimal
standard model}

\vspace{3cm}
{\large\bf
Dmitri Diakonov$^*$\footnote{
\noindent
diakonov@lnpi.spb.su},
Maxim Polyakov$^*$, Pavel Pobylitsa$^*$, \\
Peter Sieber$^\diamond$, J\"org Schaldach$^\diamond$
and Klaus Goeke$^\diamond$\footnote{
\noindent
goeke@hadron.tp2.ruhr-uni-bochum.de}} \\
\vspace{20 pt}
\noindent
{\small\it $^*$St.~Petersburg
Nuclear Physics Institute, Gatchina, St.Petersburg 188350, Russia \\
$^\diamond$Inst. f\"ur Theor. Physik II, Ruhr-Universit\"at Bochum,
D-44780 Bochum, Germany}
\end{center}
\vspace{1cm}
\abstract{The baryon number dissipation rate due to sphaleron
transitions at high temperatures in the minimal standard model is
evaluated. We find that this rate can be considerably suppressed by one
loop contributions of bosonic and fermionic fluctuations which are
particularly important for a small mass of the Higgs boson and a large
top quark mass. Fixing the latter to its recently stated value of 174 GeV
the complete erasure of the
baryon asymmetry is prevented within the framework
of the minimal standard model if the Higgs mass is less than about 66
GeV.}

\newpage

\noindent
{\bf 1.} The potential energy of the
$SU(2)$ Yang--Mills gauge field and the
Higgs field in the minimal standard model
is periodic in
the Chern--Simons number $N_{CS}$ \cite{Fadeev,Jackiw}. The vacua (with
integer $N_{CS}$) are separated by the sphaleron barrier
\cite{Dashen,Klinkhamer}
whose height is of the order of $m_W/\alpha$, where $m_W$ is the
$W$-boson mass and $\alpha=g^2/(4\pi)$ is the \SU gauge coupling
constant.

Transitions from one vacuum to a topologically distinct one over
this barrier cause a change in the baryon and lepton number by one
unit per fermion family due to the axial anomaly \cite{tHooft}. Hence,
this transition is a baryon and lepton number violating process.
Although it is
very strongly suppressed under ordinary conditions \cite{tHooft}, its rate
can become large at
high densities \cite{Rubakov,rub},
high temperatures \cite{Kuzmin,Boch,Arnold} or maybe, at high particle
energies \cite{Ringwald}.
In particular, the transitions should have occurred in the early
universe immediately after the electroweak phase transition.
Whatever was the mechanism leading to the baryon asymmetry of the universe
at earlier times, these transitions might have washed out any initial
excess of baryons over antibaryons if their rate was large enough
\cite{Kuzmin,Boch,Shap}.
In order to understand the presently observed baryon excess we therefore
need an exact determination of the transition rate.

The dominant contribution to this rate is given by the classical
Boltzmann factor $e^{-E_{\rm class}/T}$ where $T$ is the temperature, and
$E_{\rm class}$ is the classical energy of the sphaleron.
Quantum corrections arise from bosonic and fermionic fluctuations about
the sphaleron; there are also prefactors due to negative and
zero bosonic fluctuation modes \cite{Arnold}.
In \cite{Arnold} the rate was calculated
considering the classical contribution and the
factors due to zero and negative modes
only; in this case the rate is so large that any initial baryon excess
could easily have been washed out. Hence it is necessary to take the
loop corrections into account if one wants to have a chance to preserve
the asymmetry. In \cite{Boch,Shap} boson fluctuations were
considered through an effective Higgs potential resulting in an upper
limit of 45 to 55 GeV for $m_H$.

A direct evaluation of the boson determinant over non-zero modes
was made in \cite{McLCar} where an approximation method \cite{DPY}
was employed;
exact calculations were performed in \cite{McLetal,Baacke}. All those
calculations were done in the limit of high temperature in which the
four dimensional boson fluctuation matrix can be replaced by the three
dimensional one while fermions decouple completely. Although
parametrically this limit is reasonable, numerically it might
not necessarily be justified. In fact we find that numerical results
are significantly influenced by terms which vanish in the high $T$ limit,
especially by the contribution of the fermion fluctuations. Moreover, a
third independent calculation of the boson determinant is necessary
since the results of \cite{McLetal} and \cite{Baacke} deviate from each
other.

We have consistently evaluated both the boson and the fermion one loop
contributions
by the determination of the complete (discretized) spectrum of the
fluctuation matrices. This technique allows us not only to check the
existing results in the high $T$ limit but also to perform the
generalization to arbitrary temperatures. A detailed description of this
technique for the fermions can be found in \cite{rub}, an extended
paper
about the boson fluctuations is in preparation \cite{rub2}.

\bigskip\noindent
{\bf 2.} We consider the minimal version of the standard electroweak theory
with one Higgs doublet which is Yukawa coupled to left handed
fermion doublets and to right handed singlets; in the following we
write only one doublet and one pair of singlets for brevity. We neglect
the Weinberg angle, i.e.~we work with a pure $SU(2)$ gauge theory.
This idealization does not seem to be significant \cite{Kunz}.
The Lagrangian is thus
\begin{eqnarray}
{\cal L} &=&
 -\frac{1}{4g^2}F_{\mu\nu}^a F^{a\,\mu\nu}
+ (D_\mu \Phi)^\dagger
(D^\mu \Phi) - \frac{\lambda^2}{2}\Bigl(\Phi^\dagger \Phi
-\frac{v^2}{2}\Bigr)^2 \nonumber \\
 & &{}+\bar{\psi}_L i\gamma^\mu D_\mu \psi_L +
      \bar{\chi}_R i\gamma^\mu \partial_\mu \chi_R
- \bar{\psi}_L M \chi_R - \bar{\chi}_R M^\dagger\psi_L\quad\quad
\la{Lagra}\end{eqnarray}
with $F_{\mu\nu}^a=\partial_\mu A_\nu^a - \partial_\nu A_\mu^a
      + \e^{abc}A_\mu^b A_\nu^c$
and the covariant derivative being defined as
$D_\mu=\partial_\mu -\frac{i}{2} A_\mu^a\tau^a$.
$M$ is a $2 \times 2$ matrix
composed of the Higgs field components $\Phi={\phi^+ \choose \phi^0}$
and the Yukawa couplings $h_u, h_d$:
\beq
M=\matr{h_u \phi^{0\ast}}{h_d \phi^+}{- h_u \phi^{+\ast}}{h_d \phi^0}.
\la{MM}\eeq
$\psi_L$ means the \SU fermion
doublet $u_L\choose  d_L$, and with $\chi_R$ we
denote the pair of the singlets $u_R$, $d_R$. To preserve spherical symmetry
we take equal masses for the two kinds of fermions in each doublet,
i.e.~$h_u=h_d=h$. This approximation is justified for the light doublets,
but not so good for the bottom and top quark doublets. Here the mass
difference could be treated as a perturbation. However, preliminary
estimates show that the final conclusions of the present paper
do not change noticably.

The transition rate per volume $V$
of the system going from one vacuum into
a topologically distinct one is given by
the semi-classical Langer--Affleck
formula \cite{Langer,Affleck,Khleb} which,
applied to the model \cite{Arnold}, reads
\beq
\gamma=\frac{\Gamma}{V}
= \frac{\omega_-}{2\pi}\,\frac{{\cal N}_{0,-}}{V}\,\kappa_{\rm bos}\,
\kappa_{\rm ferm}\, \exp(-E_{\rm class}/T)\, ,
\la{LaAf}\eeq
with $\omega_-$ being the frequency of the
negative bosonic mode and ${\cal N}_{0,-}$ the volume factor
due to the negative and the zero bosonic modes.
$E_{\rm class}$ is the classical energy of the sphaleron configuration,
\beq
E_{\rm class}=\int d^3{\bf r}\,
\biggl[\frac{1}{4g^2}(F^a_{ij})^2+(D_i\Phi)^\dagger(D_i\Phi)
+\frac{\lambda^2}{2}\Bigl(\Phi^\dagger \Phi-\frac{v^2}{2}\Bigr)^2\biggr]=
{\cal O}\left(\frac{m_W}{\alpha}\right)\;.
\la{Eclass} \eeq
The determinants $\kappa_{\rm bos}$,
$\kappa_{\rm ferm}$ correspond to bosonic and fermionic fluctuations;
they are divergent and have to be
renormalized which we perform using a proper
time regularization scheme. The divergent parts are combined with the
classical energy, which yields the
physical parameters normalized at the scale of $m_W$.
The major part of their
temperature dependent contribution
can also be absorbed by the classical energy,
leading to temperature dependent masses (see e.g.~\cite{Kirz}):
\beq
\frac{m_W(T)}{m_W}=\frac{m_H(T)}{m_H}=\frac{m_F(T)}{m_F}
=\frac{E_{\rm class}(T)}{E_{\rm class}}
=q(T)\equiv\sqrt{1-\frac{T^2}{T_c^2}}\, ,
\la{massren}\eeq
where the critical temperature of the phase transition is
given by \cite{Kirz,rub}
\beq
T_c=\frac{1}{g}\sqrt{\frac{24\,m_W^2\,m_H^2}
{3\,m_H^2+9\,m_W^2+4\sum_{\rm doubl.}m_F^2}}\, .
\la{Tcrit}\eeq
In a realistic model all fermion masses
except the top mass are negligible, so
that $\sum_{\rm doubl.}m_F^2=\frac{3}{2}m_t^2$.
After performing this renormalization we
find
\beq
\frac{{\cal N}_{0,-}}{V}=\frac{4\pi^2 m_W(T)^8}{g^6 T^4 m_W}
\left[\sin\left(\frac{\omega_-q(T)}{2T}\right)\right]^{-1}
(N_{\rm rot}N_{\rm trans})^3\, ,
\eeq
with $N_{\rm rot}$, $N_{\rm trans}$ being the Jacobians
of the zero modes.
The determinants can be written in the following form:
\begin{eqnarray}
\kappa_{\rm ferm} &=& \frac{\prod_n{\rm cosh}\left(\frac{\epsilon_n
 q(T)}{2T}\right)}
      {\prod_n{\rm cosh}\left(\frac{\epsilon_n^0
   q(T)}{2T}\right)}
 = \exp\Biggl\{\sum_n\frac{q(T)}{2T}(|\epsilon_n|-|\epsilon_n^0|)
  \nonumber \\
 &\phantom{=}& +\sum_n\left[\ln(1+e^{-q(T)|\epsilon_n|/T})
   -\ln(1+e^{-q(T)|\epsilon_n^0|/T})\right]\Biggr\}\, ,\la{detfer}\\
\kappa_{\rm bos} &=& \left(\frac{2\,T}{m_W(T)}\right)^7\,
        \frac{\prod_n{\rm sinh}\left(\frac{\omega_n^0 q(T)}{2T}\right)}
      {\prod_n''{\rm sinh}\left(\frac{\omega_n q(T)}{2T}\right)}\;
    \frac{\prod_n{\rm sinh}\left(\frac{\omega_n^{\rm FP} q(T)}{2T}\right)}
    {\prod_n{\rm sinh}\left(\frac{\omega_n^{{\rm FP},0} q(T)}{2T}\right)}
       \quad\quad\quad \nonumber \\
&=& \left(\frac{T}{m_W(T)}\right)^7\,\exp\Biggl\{-\frac{q(T)}{2T}
    \left({\sum_n}''\omega_n - \sum_n\omega_n^0-\sum_n\omega_n^{\rm FP}
     +\sum_n\omega_n^{{\rm FP},0}\right) \nonumber \\
     &\phantom{=}& \quad -{\sum_n}''\ln\left(1-e^{-q(T)\omega_n/T}
                                     \right)
     +\sum_n\ln\left(1-e^{-q(T)\omega_n^0/T}\right)\nonumber \\
     &\phantom{=}& \quad +\sum_n\ln\left(1-e^{-q(T)\omega_n^{\rm FP}/T}
                                     \right)
     -\sum_n\ln\left(1-e^{-q(T)\omega_n^{{\rm FP},0}/T}\right)\Biggr\}
     \, .   \la{detbos}
\end{eqnarray}
Here $\epsilon_n$ are the eigenvalues of the fermionic fluctuation operator
$\delta^2 S_{\rm eff}/\delta\psi\delta\bar{\psi}$ \cite{rub},
$\omega_n^2$ are the eigenvalues of the bosonic fluctuation operator, using
the $R_{\xi=1}$ background gauge for the fluctuations \cite{McLCar,Baacke},
and $(\omega_n^{\rm FP})^2$ are the corresponding eigenvalues
of the Faddeev-Popov operator. For numerical purposes the spectra are
discretized and made finite using suitable box parameters which have
been taken large enough to ensure stability.
The double prime $\prod''$, $\sum''$
means that negative and zero mode frequencies are removed. Knowing the
eigenvalues $\epsilon_n$, $\omega_n$ and $\omega_n^{\rm FP}$ the above
determinants \eqs{detfer}{detbos} can be evaluated for any temperature
$T$. The first terms in the exponents correspond to the zero temperature
fluctuation energy while the logarithms are only present for non zero
temperature. In previous works \cite{McLCar,McLetal,Baacke} only the
high $T$ limit was considered, in which the determinants reduce to
\begin{eqnarray}
\kappa_{\rm ferm} &=& 1\, , \nonumber \\
\kappa_{\rm bos} &=& \left(\frac{1}{m_W^7}\right)\,
        \left(\frac{\prod_n\omega_n^0}{\prod_n''\omega_n}\right)\;
 \left(\frac{\prod_n\omega_n^{\rm FP}}{\prod_n\omega_n^{{\rm FP},0}}
 \right)\, .           \la{hTlim}
\end{eqnarray}
We assume the hedgehog ansatz
\[
A_i^a({\bf r}) = \e_{aij}n_j\frac{1-A(r)}{r}
+ (\delta_{ai}-n_an_i)\frac{B(r)}{r}
+ n_an_i\frac{C(r)}{r}\,,
\]
\beq
\Phi({\bf r})
= \frac{v}{\sqrt{2}}\left[H(r)+i G(r)\,{\bf n}\cdot\boldtau\right]
{\textstyle{0 \choose 1}}\;,
\la{HH}\eeq
for the classical fields and solve the corresponding equations of motion
for the profile functions $A(r),\ldots,G(r)$. Then
the fluctuation operators become block diagonal with respect to
some ''grand spin'' $K$
and can be numerically diagonalized for each $K$ separately.
More details can be found in
\cite{rub,rub2}. Hence it is possible to evaluate the
rate $\gamma(T)$ numerically
for given values of $m_H/m_W$ and $m_t/m_W$ as a function
of the temperature.

Before we present numerical results we would like to get a feeling for
the dependence of the determinants on the mass parameters. Though the
exact numerics is quite complicated, it is easy to find the behaviour of
the determinants for large top quark mass and for small Higgs mass. For
large top quark mass the square loop diagram in the external Higgs field
is the dominant
contribution, and the fermion sea energy is proportional to
$N_c(h\phi)^4\ln(h\phi/m_W)$, where $h$ is the Yukawa coupling and
$\phi$ is the Higgs field of the sphaleron. To get the energy
one has to integrate this expression over the range of space
where the Higgs field differs from its vacuum expectation value,
i.e.~over the spread of the sphaleron. For large
spatial distance $r$ the asymptotics of the fields is dominated by the
term $e^{-m_Hr}$ so that the sphaleron size is roughly
$\sim m_H^{-1}$ if $m_H<m_W$. Hence all matrix elements
of the fluctuation matrices behave as $m_H^{-3}$. This scaling law
holds for all contributions to the determinant; i.e.~for small
Higgs masses the fluctuations show a strong increase. It is this
increase which is finally responsible for the suppression of the baryon
number dissipation and provides an upper limit for $m_H$ to ensure the
preservation of the asymmetry. Thus for $m_H<m_W<m_t$ we can estimate
\begin{eqnarray}
\ln(\kappa_{\rm ferm}) &\sim& N_c\, \frac{(m_t/m_W)^4\ln(m_t/m_W)}
{(m_H/m_W)^3}\, , \la{ferapp} \\
\ln(\kappa_{\rm bos}) &\sim& \frac{1}{(m_H/m_W)^3}\, . \la{bosapp}
\end{eqnarray}
These are the quantum corrections to the
classical sphaleron energy $E_{\rm class}$ \ur{Eclass}. Though
parametrically they are $\alpha$ times
smaller than $E_{\rm class}\,$,  numerically the fermion sea
contribution to the sphaleron energy
appears to be large, especially for large top masses $m_t$ and
relatively small Higgs masses $m_H$.
It is mainly the fermionic factor $\kappa_{\rm ferm}\,$,
which was put to unity in the previous work
\cite{Boch,McLCar,McLetal,Baacke}, that leads to a significant
additional suppression of the baryon dissipation rate, see below.

\bigskip\noindent
{\bf 3.} Sphaleron transitions can increase and decrease the baryon number.
If the baryon number $B$ were zero,
the transitions in both directions would happen equally
often and cancel each other. In the
case $B\ne 0$ one has to introduce a chemical potential;
it favours transitions which erase
the baryon asymmetry in accordance with the le Ch\^atelier principle.
This has been done in \cite{Boch,Arnold},
and for fermions with small masses one gets:
\beq
\frac{1}{B}\,\frac{dB}{dt}=-\frac{13}{2}\,\frac{\gamma(T)}{T^3}\, .
\eeq
Since the top mass is actually not
small the prefactor $\frac{13}{2}$ should be replaced by a slightly
bigger number. This effect, however,
is negligible compared to the other factors so we
do not consider it further.

In this letter we assume, in accordance with the standard model,
that $B-L$ is conserved; we do not consider a possibility that
there might be a primodial excess of say, antileptons owing to
unknown forces violating the $B-L$ number. In that case the
sphaleron transitions would, on the contrary, lead to the yield
of baryons.

Standard cosmology gives a relation
between time and temperature \cite{Weinberg}:
\beq
t=C\,T^{-2}\, ,
\eeq
with the constant $C\approx 5\cdot 10^{15}\, m_W$
depending on the Planck mass and
the number of degrees of freedom of the thermalized particles.
Hence we obtain
\beq
\frac{1}{B}\,\frac{dB}{dT}=13\, C\,\frac{\gamma(T)}{T^6}\, ,
\eeq
which can be integrated to
\begin{eqnarray}
B(T)&=&B(T_c)\,\exp\left\{-13\,C\int_T^{T_c}
\frac{\gamma(T)}{T^6}\,dT\right\}\nonumber \\
    &=&B(T_c)\,\exp\left\{\frac{-13\,C}{T_c^5}\int_0^{q(T)}\frac{q\,\gamma(q)}
       {(1-q^2)^{7/2}}\,dq\right\}\;,
\la{Bquot}
\end{eqnarray}
where $q(T)$ was defined in \eq{massren}.
Thus, we can evaluate the ratio $B(0)/B(T_c)$ of the
present baryon number, $B(0)$, to the
one immediately after the phase transition, $B(T_c)$. From this ratio
we shall conclude on an upper bound of the Higgs mass.

\bigskip\noindent
{\bf 4.} For our numerical calculations
we fixed the coupling constant to its physical value
$g=0.67$, $m_W=$ 83 GeV,
and the top quark mass to $m_t=2.1\,m_W\,$, according to its recently
stated value of 174 GeV.
The only free parameter left is the Higgs mass $m_H$.

First we recalculated $\ln(\kappa_{\rm bos})$ of \eq{hTlim}
in the high $T$ limit which was
previously done in \cite{McLetal,Baacke} with deviating results.
We find that our values are between those of the two
preceeding computations; they agree with
the ones of ref.~\cite{Baacke} up to about $10\%$. Apart from numerical
uncertainties this difference could be
probably explained by the fact that in
ref.~\cite{Baacke} the complete tadpole expansion except the term
linear in temperature was removed due to the loop renormalization,
while we have performed the renormalization at zero temperatures strictly,
as it is usually done. Our scheme corresponds to a subtraction of only the
first term in the tadpole expansion so that the difference lies
in higher order terms which vanish in the high $T$ limit but
can influence the numerical result. There is a
larger deviation from the results of \cite{McLetal}, only a qualitative
agreement is found.

A powerful check of our numerical performance was that
determinants were computed in several completely different gauges
for the sphaleron field. The fluctuation operators change drastically
when one goes from on gauge to another, however we have checked that
the gauge invariant eigenvalues change only in the range of $0.5\%$,
giving an estimate of the numerical accuracy of the spectrum.

Our aim is, however, a generalization to arbitrary temperatures.
Figure 1 shows the logarithms of
the different contributions to the transition rate
$\gamma$ of \eq{LaAf}: the
classical part $e^{-E_{\rm class}(T)/T}$, the prefactor
$\omega_-\,{\cal N}_{0,-}/(2\pi V)$, and the bosonic and fermionic
fluctuations $\kappa_{\rm bos}$ and $\kappa_{\rm ferm}$.
It is convenient to take the parameter
$q(T)$ (see \eq{massren}) as independent variable rather than the
temperature itself.
Obviously, the loop corrections, especially the
fermionic ones yield a strong suppression of the total transition rate.
If the temperature approaches the critical value of the electroweak
phase transition,
all masses disappear, and the transition rate goes rapidely to
zero due to the vanishing prefactor.
It should be mentioned that the one-loop approximation we are dealing with
breaks down in the near vicinity of the phase transition. However, the
integral in \ur{Bquot} is
strongly dominated by a small region around the maximum
of the transition rate $\gamma(q)$, which is separated from the phase
transition itself (see Figure 1).
Figure 1 was calculated for
$m_H=0.8\,m_W$, for other values of $m_H$ the qualitative
behaviour of the curves is
basically the same, only the numbers change somewhat.

According to \eq{Bquot} we calculated the
quotient $B_0/B_{T_c}$ of the present baryon
number $B_0$ and the initial one $B_{T_c}$.
Figure 2 shows the ratio for different values of $m_H$.
If we assume that the initial baryon number
$B_{T_c}$ was not larger than about $10^5\,B_0\;$
\cite{Boch}, we find an upper bound of the Higgs mass of about
$0.79\,m_W\approx 66\,$GeV.
Figure 2 exhibits a very strong dependence on $m_H$:
even if we assume an apparently unrealistic
initial baryon excess $B_{T_c}\approx 1$, we get from
the present-day asymmetry $B_0\approx 10^{-10}$
that the Higgs mass does not exceed the value of
$m_H\approx 0.81\,m_W\approx 67\,$GeV.
On the other hand, if we assume $B_{T_c}=B_0$,
the upper limit for $m_H$ lies at
$m_H\approx 0.7\,m_W\approx$ 58 GeV.

In order to emphasize the significance of the
fermionic fluctuations, which have not
been taken into account previously, we calculated
the transition rate and the
resulting decrease of the baryon number also without
the fermionic contribution.
Figure 2 shows that in this case the upper bound for
the Higgs mass,
assuming $B_{T_c}=10^5B_0$, would be as low as
about 51\,GeV which is close to the approximate
estimate of ref.~\cite{Boch,Shap}.

To summarize: For any temperature $T<T_c$ both
bosonic and fermionic fluctuations suppress the sphaleron transition
rate considerably,
depending on the value of the Higgs mass. Setting the top quark mass to its
recently claimed value of 174 GeV, the condition that the baryon
asymmetry is not washed out by sphaleron transitions leads
to an upper limit of 66 GeV for the Higgs boson mass. This numerical result
is based on the Langer--Affleck formula and applies
to one loop calculations within the minimal standard model with
one Higgs doublet. Our upper limit is compatible with the
experimental lower limit of about 60 GeV \cite{Wyatt,Sopczak}
provided we include the fermionic corrections. If no Higgs particle
is found below our upper limit, this would be a clear sign that the minimal
standard model should be extended, maybe to a model with two Higgs doublets.

\newpage\noindent
{\Large\bf Figure Captions} \par
\bigskip\noindent
{\it Figure 1:} Classical and loop contributions
to the sphaleron transition rate per volume,
$q=\sqrt{1-(T/T_c)^2}\,,\;T_c=0.957\,m_W\,,\;m_H=0.8\,m_W\,$.
The interval which actually
contributes to the integral \ur{Bquot} is marked.
\par\bigskip\noindent
{\it Figure 2:} Total decrease of the baryon number via the sphaleron
transition
in dependence of the Higgs mass, with and without fermionic
fluctuations taken into account. $B_0/B_{T_c}$ is the ratio of the
present baryon number to the one of the electroweak phase transition.
\end{document}